\documentclass[aps,pra, onecolumn,14 pts,a4paper]{revtex4}
 \pdfoutput=1

\usepackage{amssymb,amsmath,amsfonts,graphicx}
\usepackage{bm}
\usepackage{epsfig,color,times}
 \usepackage{bm,color}
\usepackage{graphicx}
\usepackage{amssymb}
\usepackage{epstopdf}

\begin{document}


\title{Recording of Leray-type singular events in a high speed wind tunnel}

\author{Martine Le Berre$^1$, Yves Pomeau$^2$ 
,} 

\affiliation{$^1$  ISMO-CNRS, Universit\'e Paris-Sud, 91405 Orsay Cedex, France.
\\$^2$  Ladhyx, Ecole Polytechnique, Palaiseau, France.
}

\date{\today }

\begin{abstract}
It has long been suspected that flows of incompressible fluids at very large or infinite Reynolds may present finite time singularities. We review briefly the theoretical situation on this point. Then we show that single point records of velocity fluctuations in the Modane wind tunnel show correlations between large velocities and large accelerations that are in agreement with the scaling laws for such singularities as derived by Leray in 1934. Conversely the experimental correlations between velocity and acceleration are not explainable by Kolmogorov scalings. This implies in particular that the singularities cannot be seen as the end of a cascade in the sense of Kolmogorov, but are best described as singular events in the sense of Leray.

\end{abstract}

\maketitle

\section{Introduction}
Much remains to be understood concerning  the "generic behavior" of solutions of the fluid equations for incompressible flows at very large Reynolds number. Kolmogorov's theory  is based on the idea that every quantity scales with the power dissipated per unit time and mass. Assuming the viscosity to be negligible at the scales of observation, one needs to introduce, to explain dissipation, a transfer of energy from large to small scales where viscosity becomes significant. This approach predicts well the spectra of velocity fluctuations as a function of the wave number \cite{frisch}.
However it was quickly realized that this approach is not able to describe other observed phenomena like intermittency \cite{interm} which can be seen as the occurrence of large velocity fluctuations not describable at all by Gaussian or quasi-Gaussian statistics. Besides the property of energy conservation of the Euler fluid equations, very little of the properties of those equations is used to derive Obukhov-Kolmogorov spectra. The present work intends to explain first the idea of self-similar solution, and then to show that some of its consequences can be directly observed in the fluctuations of velocity and acceleration recorded in the highly turbulent flow of the Modane wind tunnel.

In 1934 Leray \cite{leray} published a paper on the equations for an incompressible fluid in 3D, with and without viscosity. He introduced many important ideas, among them the notion of weak solution and also the problem to be solved to show the existence (or not) of a solution becoming singular after a finite time when starting from smooth initial data. 

Over the years this motivated a lot of works, mostly by mathematicians, the main effort being to try to prove or disprove the existence of such singularities assuming properties of the initial data. Other attempts have been directed toward a direct solution of the dynamical Euler equations, with the purpose of showing they have or not a finite time singularity. 

It has been  suggested recently  \cite{YP},  \cite{modane} 
  to find a direct numerical solution of the Leray equations for a self-similar singularity of the Euler equations (or Euler-Leray equations). 
The Euler equations read: 
\begin{equation}
\frac{\partial{\bf{u}}}{\partial t} + {\bf{u}}\cdot \nabla {\bf{u}} = - \nabla p
\textrm{,}
\label{eq:Euler1}
\end{equation}
and 
\begin{equation}
\nabla \cdot {\bf{u}} = 0 
\textrm{,}
\label{eq:Euler2}
\end{equation}

Leray looked at the Navier-Stokes equations, which amounts to add $\nu \nabla^2  {\bf{u}}$ to the right-hand side of equation (\ref{eq:Euler1}). Specifically he looked at solutions of the self-similar type: 
\begin{equation}
{\bf{u}}( {\bf{r}}, t) = (t^*- t)^{-\alpha}  {\bf{U}} ( {\bf{r}}(t^*- t)^{-\beta})\textrm{,}
\label{eq:self}
\end{equation}

 where $t^*$ is the time of the singularity (set to zero afterwards), where $\alpha$ and $\beta$ are positive exponents to be found and where  the field with upper-case letters $ {\bf{U}}(.)$ is to be derived by solving Euler, or Navier-Stokes equations.

That such a velocity field is a solution of Euler or Navier-Stokes equations implies 
\begin{equation}
  \alpha + \beta =1
 \textrm{,}
 \label{eq:alpha1}
\end{equation}
 a  condition which ensures the balance between the two terms in the l.h.s. of (\ref{eq:Euler1}), which are respectively of order $t^{-(\alpha+1)}$  and $t^{-(2\alpha+\beta)}$.

Below we shall compare our predictions with experimental data  taken from the Modane wind tunnel,  where the velocity  field is advected by a mean flow. In this case  the self-similar solution (\ref{eq:self}) describes  the  behavior of the fluctuations  of the velocity field,  $ {\bf{\delta v}}= {\bf{v}}- <{\bf{v}}>.$   But  the  balance condition (\ref{eq:alpha1})  is still valid, because  the advection term, $< {\bf{v}}>\cdot \nabla {\bf{\delta v }}$, coming from the mean flow, of  order $t^{-(\alpha+\beta)}$,  is smaller  than the others two.  

In the case of Navier-Stokes equation, the balance with the dissipative term  $\nu \nabla^2  {\bf{u}}$, of order  $t^{-(\alpha+2\beta)}$, imposes $\beta=1/2$,   which yields  the exponents found by Leray for the case of the  Navier-Stokes equations

\begin{equation}
 \alpha = \beta =1/2
 \textrm{,}
 \label{eq:alpha2}
\end{equation}

\section{ Euler-Leray's equations}

In  the case of Euler equation, there are several possibilities to get a second relation between the two exponents, according to what conservation laws  are considered.    Let consider first the  constraint of conservation of circulation.  The circulation $\Gamma=\oint{u ds }$ along a closed curve carried by the flow toward  the singularity, is of order $U^{2 }t^{\beta-\alpha}$. Then the condition for conservation of circulation implies $ \alpha - \beta=0$, that gives (\ref{eq:alpha2}), namely the  same exponents as for  Navier-Stokes case.  
Moreover the velocity scales like 
\begin{equation}
u(r, t) \sim (-t)^{-1/2} \Gamma^{1/2}
 \textrm{,}
 \label{eq:circ}
\end{equation}
near the singularity because $\Gamma \sim U^{2}$. With such a choice, the  total energy of solutions of the self-similar problem is diverging, but the divergence of the energy  does not imply the absence of singularity  \cite{modane}. 

If one imposes instead of the conservation of circulation that the energy in the collapsing domain is conserved, one must satisfy the constraint $- 2 \alpha + 3 \beta = 0$, which yields $\alpha = 3/5$ and  $\beta = 2/5$, the Sedov-Taylor  exponents. 
No set of singularity exponents can satisfy both constraints of energy conservation and of constant circulation on carried closed curves.

We choose $\alpha = \beta = 1/2$ in the following, or
\begin{equation}
{\bf{u}}( {\bf{r}}, t) = (t^*- t)^{-\frac{1}{2}}  {\bf{U}} ( {\bf{r}}(t^*- t)^{-\frac{1}{2}})
 \textrm{,}
 \label{eq:self2}
\end{equation}

With upper case letters for the position, 
$ {\bf{R}} =  {\bf{r}}(- t)^{-1/2}$, the Euler equations become the Euler-Leray equations for $ {\bf{U}}({\bf{R}})$,
\begin{equation}
\frac{1}{2}({\bf{U}}  + {\bf{R}}\cdot \nabla {\bf{U}}) + {\bf{U}}\cdot \nabla {\bf{U}} = - \nabla P
\textrm{,}
 \label{eq:EulerU}
\end{equation}

and 
\begin{equation}
 \nabla \cdot {\bf{U}} = 0
 \textrm{.}
 \label{eq:DivU}
\end{equation}

A singularity of the self similar type must decay at large distances in such a way that, at such large distances (in the stretched variables),  it becomes independent on time. Otherwise it would depend singularly on time everywhere and so not be a point wise singularity. Moreover the solution of  the Euler-Leray equations must be smooth as a function of ${\bf{R}}$. Otherwise it makes a singular solution at any time, not at a single time, like for example solutions of the type of Landau submerged jet \cite{landau jets} which are singular uniformly in time, and cannot belong to the class of solutions considered here. 

The first constraint (solution independent on time at large distances) is satisfied if ${\bf{U}} \sim 1/R $ at $R$ large. Returning to the initial space-time dependence one gets the asymptotic behavior (in the stretched variable) $ u  \sim  (-t)^{-1/2}/ r (-t)^{-1/2}  \sim 1/r$ with no time dependence.  At $t =0$ (time of singularity) the velocity field of the singular solution is exactly like $1/r$ times a function of the angle to satisfy incompressibility (a property perhaps experimentally checkable by particle image velocimetry). 

The $1/R$ behavior of the solution cancels the linear term in Euler-Leray (and NS-Leray as well),  dominant at large distances, as it should. This leads to seek a formal Laurent expansion of this solution in inverse powers of $R$, 
\begin{equation}
 {\bf{U}}({\bf{R}}) =  \sum_{n =1}^{\infty} \frac{1}{R^n} {{\bf{W}}_n}  (\hat{\bf{R}})
\textrm{.}
\label{eq:Euler2exp}
\end{equation}
with $\hat{\bf{R}} = {\bf{R}} /R$ unit vector. 
Putting this expansion of ${\bf{U}}$ into Euler-Leray, we get
\begin{equation}
\frac{1}{2} \sum_{n = 3}^{\infty}(1 - n)  \frac{1}{R^n} {{\bf{W}}_n}  (\hat{\bf{R}}) + {\bf{U}}\cdot \nabla {\bf{U}} = - \nabla P 
\textrm{.}
\label{eq:Pom4}
\end{equation}
Knowing ${\bf{W}}_1(\hat{\bf{R}})$ (arbitrary at this step) this can be mapped in an iteration for computing ${\bf{W}}_3(\hat{\bf{R}})$, ${\bf{W}}_5(\hat{\bf{R}})$, etc. The pressure being derived from the incompressibility condition. This is practically very cumbersome. Other methods have to be found to solve Euler-Leray, as explained now: the idea is to replace the Euler-Leray equation by an iteration.  

Define the $k$- Cartesian component of the nonlinear part of Euler-Leray as, 
\begin{equation}
 V_k({\bf{R}}) =  \partial_k  P +  U_j \partial_j U_k
\textrm{.}
\label{eq:EulerV}
\end{equation}
If one assumes the vector ${\bf{V}}$ to be known one can solve formally Euler-Leray by integration on the modulus of ${\bf{R}}$, 
\begin{equation}
 U_k({\bf{R}}) = \frac{1}{R}   \int_0^{R}  {\mathrm{d}}R' (- 2 V_k(R', {\hat{\bf{R}}})) 
 \textrm{.}
\label{eq:Euler6}
\end{equation}

The basic principle for an iterative solution is to assume that the left-hand side is known, put the rest in the (non linear) right-hand side, compute this right-hand side for a given field satisfying basic constraint. This yields an estimate for the velocity field which can be put into the right-hand side and the iteration is  continued in principle until it converges to the desired fixed point which is the solution of the equations one started from. However things do not work this well for a number of reasons.  This runs into a number of difficulties that have yet not been got rid out.  There is a fundamental point that gives some optimism for the existence of such a non trivial solution: the Euler-Leray equation has a variational formulation \cite{YP}, as the original Euler equations, and it is likely that a non trivial extremum of the corresponding functional exists. 

To conclude on the Euler-Leray equation, it yields a well defined schema for the existence of solutions of the Euler equations in 3D, becoming singular in a finite time at a single point. A by-product of this analysis is the set of exponents of the singularity which may be compared with the experimental data for the large fluctuations observed in the records of time dependent velocity in a turbulent flow, as done below.

One obvious motivation for working on Euler-Leray singularities is their possible connection with the (loosely defined) phenomenon of intermittency in high Reynolds number flows. This raises several question: 

1. What is the difference between Euler-Leray and Navier-Stokes-Leray singularities? 

2. What is specific to Leray singularities compared to other schema for intermittency? 

3. What would be specific of an Euler-Leray singularity in a time record of large Reynolds number flow?

Point 1: Difference between Euler-Leray and Navier-Stokes-Leray: little is known about it, in particular do both have nontrivial solutions, or does none has nontrivial solutions or only one has nontrivial solution? 

Point 2 : If intermittency is caused by Leray-like singularities, they should show a strong positive correlation between singularities of the velocity and the acceleration (see below). Compared to predictions derived from Kolmogorov theory this (positive) correlation is a strong indication of the occurrence of Leray-like singularities near large fluctuations.  It is fair to say however that Kolmogorov himself never mentions this question of finite time singularity of either Navier-Stokes  or Euler equations. So it would be unfair to attribute to him any claim about those singularities. 

Point 3: Both in Euler-Leray and NS-Leray cases, the velocity field at the singular time scales like $1/r$, $r$ distance to the singularity. 

The scaling laws derived for the velocity-acceleration correlations in time records of velocity is fairly simple. First  in the case of  Euler equation,  the order of magnitude of  the velocity in terms of the circulation close to the singular point, equation (\ref{eq:circ}), is associated to the relation  $U_{Euler} \sim \Gamma^{1/2}$, as written  above.

 In the case of the Navier-Stokes equation  the solutions  depend only on one dimensionless number, which is locally  (close to the singularity) of order 
  \begin{equation} 
  Re_{s} \sim  \frac{\Gamma}{\nu}
   \textrm{.}
  \label{eq:Re}
\end{equation}
The dissipation imposes  therefore to multiply the  pre-factor $\Gamma^{1/2}$  by a  numerical function   $f(Re_{s})$  of the Reynolds number. One  
 can write $U_{N-S}\sim \Gamma^{1/2}  f(\Gamma/\nu) $,   
 that gives the  relation,
  \begin{equation} 
u(r, t) \sim (-t)^{-1/2} \Gamma^{1/2} f(\Gamma/\nu)
 \textrm{.}
 \label{eq:circNS}
\end{equation}
The numerical value of $f$ depends on the precise solution we are considering, and of the value of the extension of the path  $r(-t)^{(-1/2}$ defining the circulation.  
 As the kinematic viscosity of the fluid, $\nu$  tends to zero  at fixed $\Gamma$ (or $Re_{s}$ tends to infinity), $f$  tends to the limit value  $f(\infty)$ which should be a finite number.  Because this limit correspond to the case of Euler equation, assuming that the solutions (for Euler and N-S equations) merge, we get  $f(\infty)=1$,
which gives  the  order of magnitude of a self-similar solution  in N-S  case in this limit.   But for finite values of the Reynolds number $Re_{s}$, the relation (\ref{eq:circNS})  cannot give a direct estimate of the magnitude of N-S self-similar solution  because the function $f()$ is unknown at this time.  In that case the local Reynolds number cannot be deduced from  experimental results,  see below in the next section.

From the  condition of conservation of  the circulation  close to the singularity,  and  (\ref{eq:Re}),  we deduce that  the  local Reynolds number  is also constant (in order in magnitude) in the collapsing domain, because the constraint on of the circulation imposes a balance between   the  growth of the velocity  and the stretching of the  singular domain. This property contradicts the standard ideas on turbulence  according to which the only relevant parameter is the power dissipated per unit mass and time.
 Here we find that in such singular events, even though the length scales become very small, the Reynolds numbers typical of those small domains do not tend to zero, but instead stay constant, because the  velocity grows continuously (at the same pace) as the space scale decreases. 

Let us introduce the \textit{effective} circulation
\begin{equation}
\tilde{\Gamma}^{1/2} = \Gamma^{1/2} f(\Gamma/\nu)
 \textrm{.} 
 \label{eq:effcirc}
\end{equation}
for a self-similar solution in  the Euler and N-S cases, for simplicity.  We can  immediately  derive the scaling for the acceleration $\gamma(r, t)$,  that is $\gamma(r, t) \sim (-t)^{-3/2} \tilde{\Gamma}^{1/2}  $.  Accordingly one finds the time independent relation 
\begin{equation}
u^3  \sim  \gamma  \tilde{\Gamma} \textrm{.} 
 \label{eq:cubv}
\end{equation}

Let us now turn to the Kolmogorov scaling. The starting point is the famous Kolmogorov relation $ u_r  \sim  (\epsilon r)^{1/3}$ where $u_r$ is the typical change of velocity over a distance $r$ and $\epsilon$ is the power dissipated per unit mass of the turbulent fluid. With those scaling and $r\sim u_{r}t$ the acceleration $\gamma\sim u_{r}/t  \sim u_{r}^{2}/r$ becomes  of order $\gamma  \sim \epsilon^{2/3} r^{-1/3}$. Therefore the relationship between $u_r$ and $\gamma$ independent on $r$ is  
\begin{equation}
\gamma u_r  \sim  \epsilon  \textrm{.} 
 \label{eq:Kolm}
\end{equation}
The latter expression is in complete contradiction with (\ref{eq:cubv} )   deduced for self-similar solutions, a problem that we  shall consider just below  by comparing with experimental data.

\section{ Experimental results }

The two  relations (\ref{eq:cubv}) and (\ref{eq:Kolm}) can be tested against the experimental results by comparing the values of the velocity fluctuations $\delta v=v-<v>$ and of the acceleration $\gamma$ recorded at the same point and the same time in the domain of large accelerations.   Our aim is to use experimental data in order to look at the occurence of self-similar solutions  in the turbulent flow, and more precisely to search if self-similar solutions of type (\ref{eq:self2}) exist actually in the flow.   If  such solutions exist, even as rare events, they should be visible at least in the large accelerations domain where one should expect to obtain a relation  of type (\ref{eq:cubv}) between acceleration and velocity fluctuations. On the contrary, if  Kolmogorov-scalings  are the only ones which drive $\it{all}$ the dynamics, large accelerations should occur  predominantly when the velocity fluctuations are close to zero. 

\begin{figure}
\centerline{ 
\includegraphics[height=2.50in]{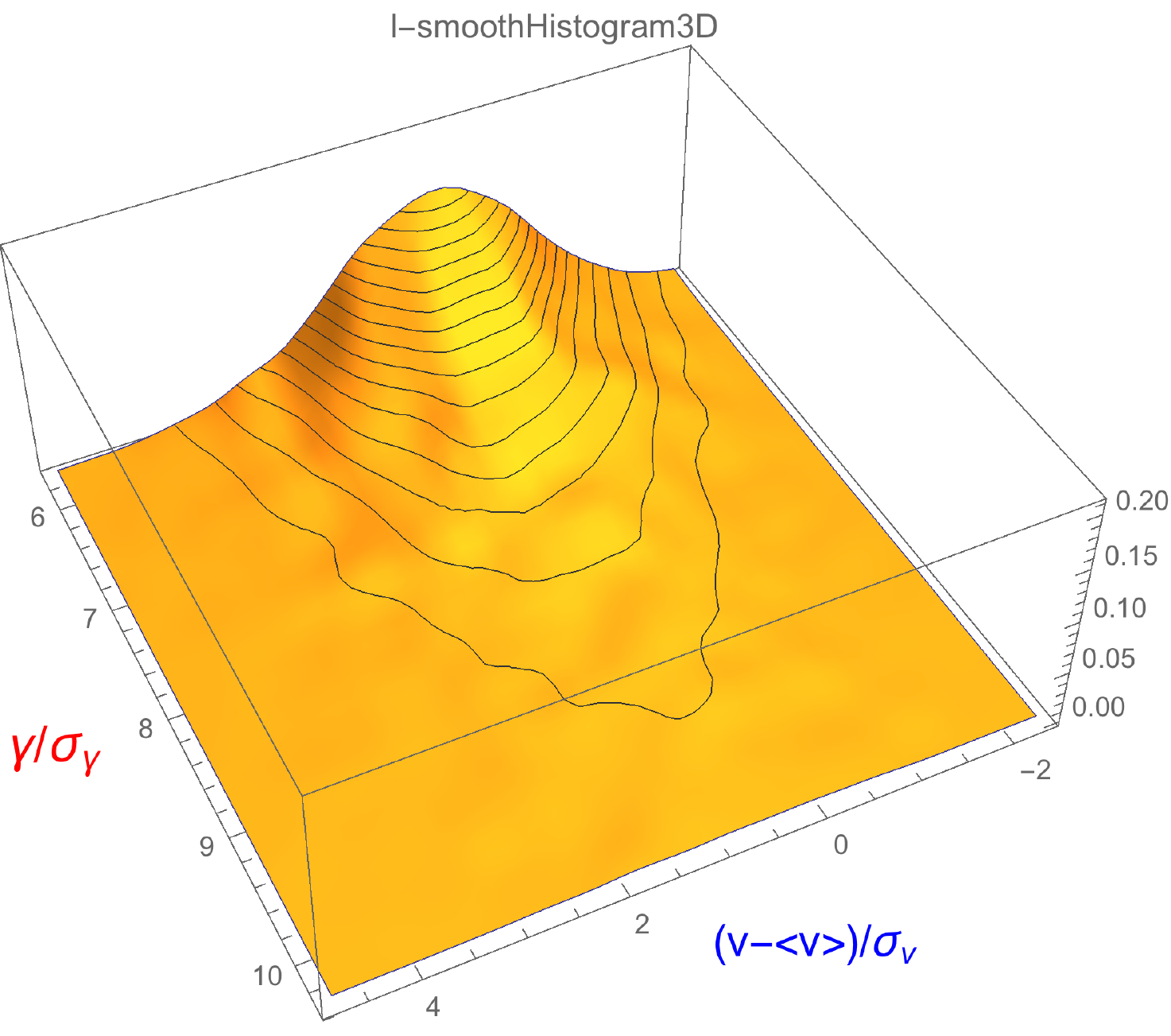}
  }
\caption{Histogram of the coupled variables $v(t_{i}) ,\gamma(t_{i})$ in the large acceleration domain, $ \gamma \ge 5\sigma_{\gamma} $. }
\label{fig:histogram}
\end{figure}

In other words, we want to see if large accelerations  occur for large  or for small velocity fluctuations.  We looked  at the data obtained in the wind tunnel of Modane, where the turbulent Eulerian velocity was recorded by a single hot wire.
Details of the experimental set-up can be found in Refs.\cite{expmod}, \cite{exp2mod}. Let us shortly recall the  conditions of this  experiment : the Reynolds number  is equal to $Re_{\lambda}=\sqrt{15Re}= 2500$, so that the regime is in fully developed turbulence. The measurements were made in the return vein of the wind tunnel, where turbulence is not really isotropic, but mainly resulting from the separation of an unstable boundary layer. The sampling time is $t_{s}= 1/25 ms$ ($f_{s} = 25 khz$). It is smaller than the dissipation time $ t_{\eta}=1/10.7 ms$.  In the following we assume ergodicity of the velocity fluctuations, it follows that any average is calculated as a temporal average, with time running over the full  data ( $13.7 \; 10^{6}$ points)
covering a total record time of about $10$ minutes. 
The average velocity is $<v>= 20.545 m/s$, and the standard deviation is $\sigma_{v}=  1.69m/s$. 
 In figure \ref{fig:histogram}  we show   the 2D-histogram of velocity fluctuation and 	acceleration recorded  at the same time, in  the domain  of large accelerations, $ \gamma \ge 5 \sigma_{\gamma} $. We show that there are more events for positive velocity fluctuations than for negative and null ones.
 This qualitative observation is in favor of the existence of self-similar solutions, but is not sufficient to claim that they are of the form  (\ref{eq:self2}).  To investigate a quantitative relationship between $\gamma$ and $(\delta v)^{n}$,  for various values of the integer $n$,  we have calculated the conditional  momentum of the velocity fluctuations $<(\delta v)^{n }>_{/\gamma}$, which are the average values of $(\delta v)^{n} $  weighted by the conditional probability  of the velocity $P_{\gamma}(\delta v)$ for a given value $\gamma$ of the acceleration,
\begin{equation}
<( \delta v)^{n}>_{\gamma}=  \int{ d(\delta v) (\delta v)^{n}     P_{\gamma}(\delta v)      }
\textrm{.} 
 \label{eq:moment}
\end{equation}
Setting $u=\delta v$ for ease, the  probability  $P(u_{i},\gamma_{j})du d\gamma$ for the set of variables $(u,\gamma)$ to be inside the domain $ [(u_{i},u_{i}+du)\cup (\gamma_{j}, \gamma_{j } + d\gamma)]$ , is given by the number of points recorded in this domain divided by the total number of recorded points, $\mathcal{N}$,
 \begin{equation}
P(u_{i},\gamma_{j}) du d\gamma=  N_{i,j} /\mathcal{N}
\textrm{.} 
 \label{eq:pij}
\end{equation}

The conditional probability $ P_{\gamma_{j}}(u_{i}) du$ for the velocity to be inside the domain $[u_{i},u_{i}+du]$ if  the acceleration is inside the domain $ [\gamma_{j}, \gamma_{j } + d\gamma] $, is  given by   $ N_{i,j} / N_{j}$, where $N_{j}=\sum_{i} N_{i,j}$. Using (\ref{eq:moment})-(\ref{eq:pij}), we get  the  following expression for the momenta in terms of the number of points recorded in the elementary domains, 
\begin{equation}
< u^{n}>_{\gamma_{j}}=  \sum_{i} u_{i}^{n} \frac {N_{i,j}} {N_{j}}
\textrm{.} 
 \label{eq:moment2}
\end{equation}

   \begin{figure}
\centerline{ 
(a)\includegraphics[height=2.2in]{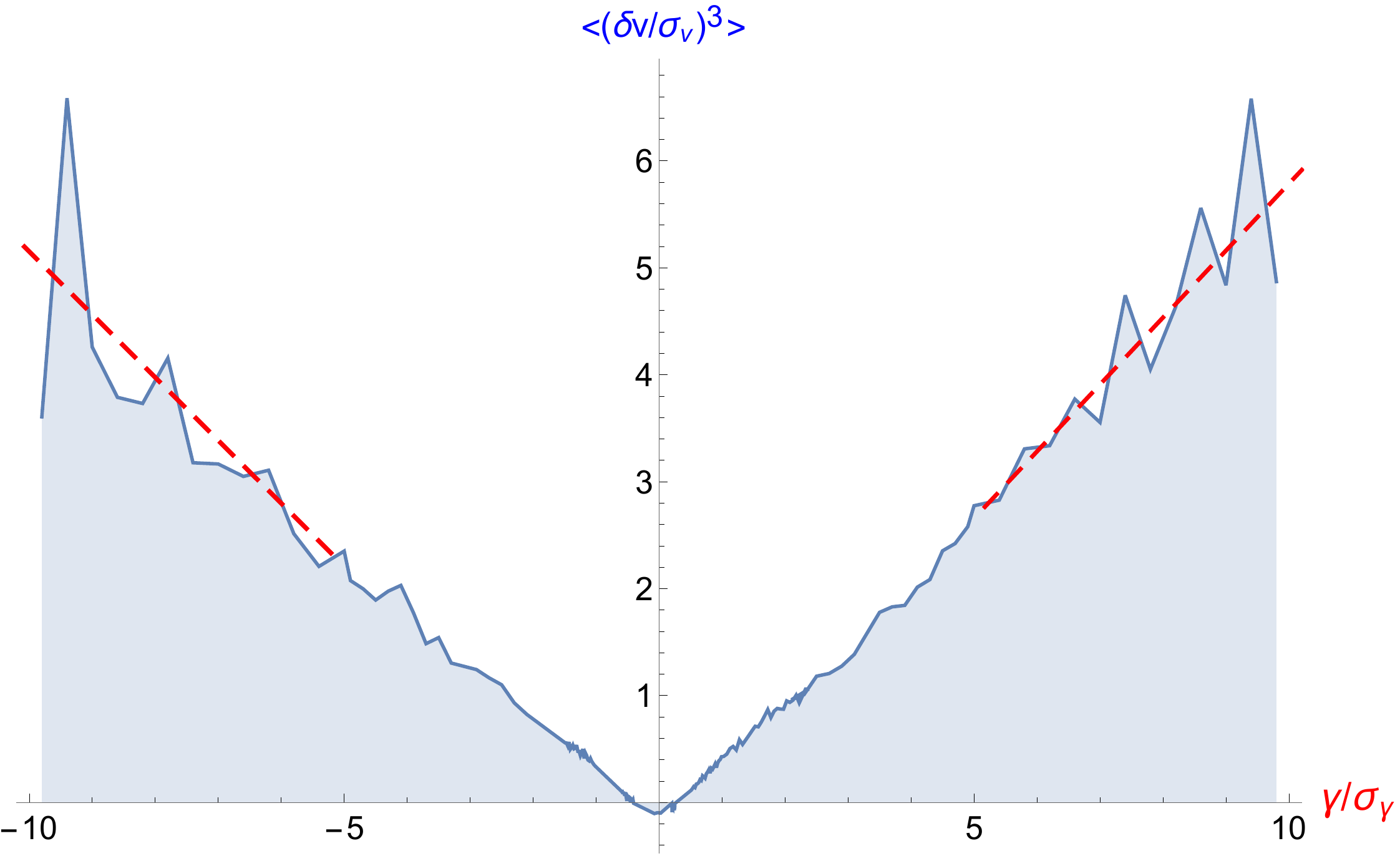}
(b)\includegraphics[height=2.2in]{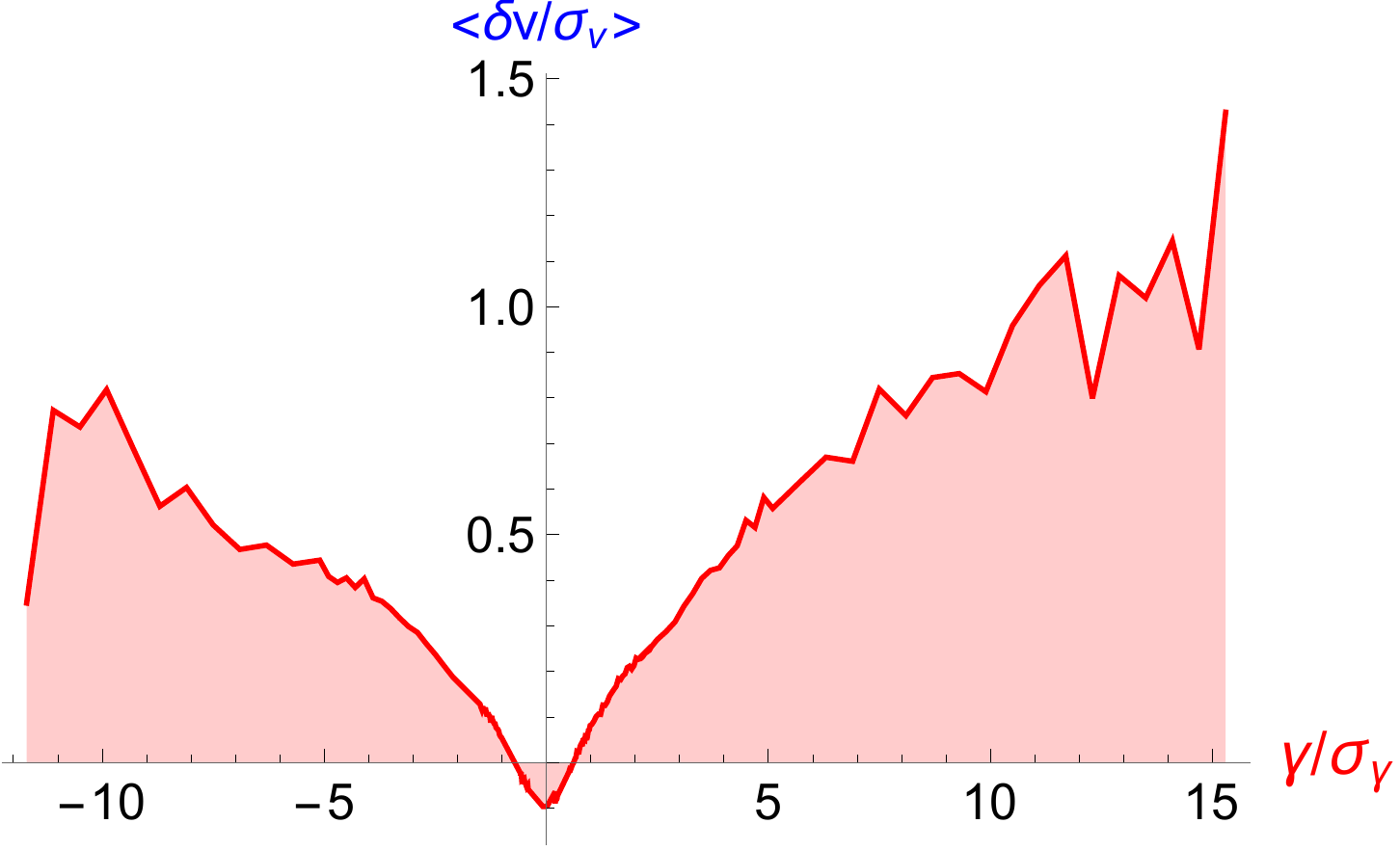}
  }
  \centerline{ 
(c)\includegraphics[height=2.0in]{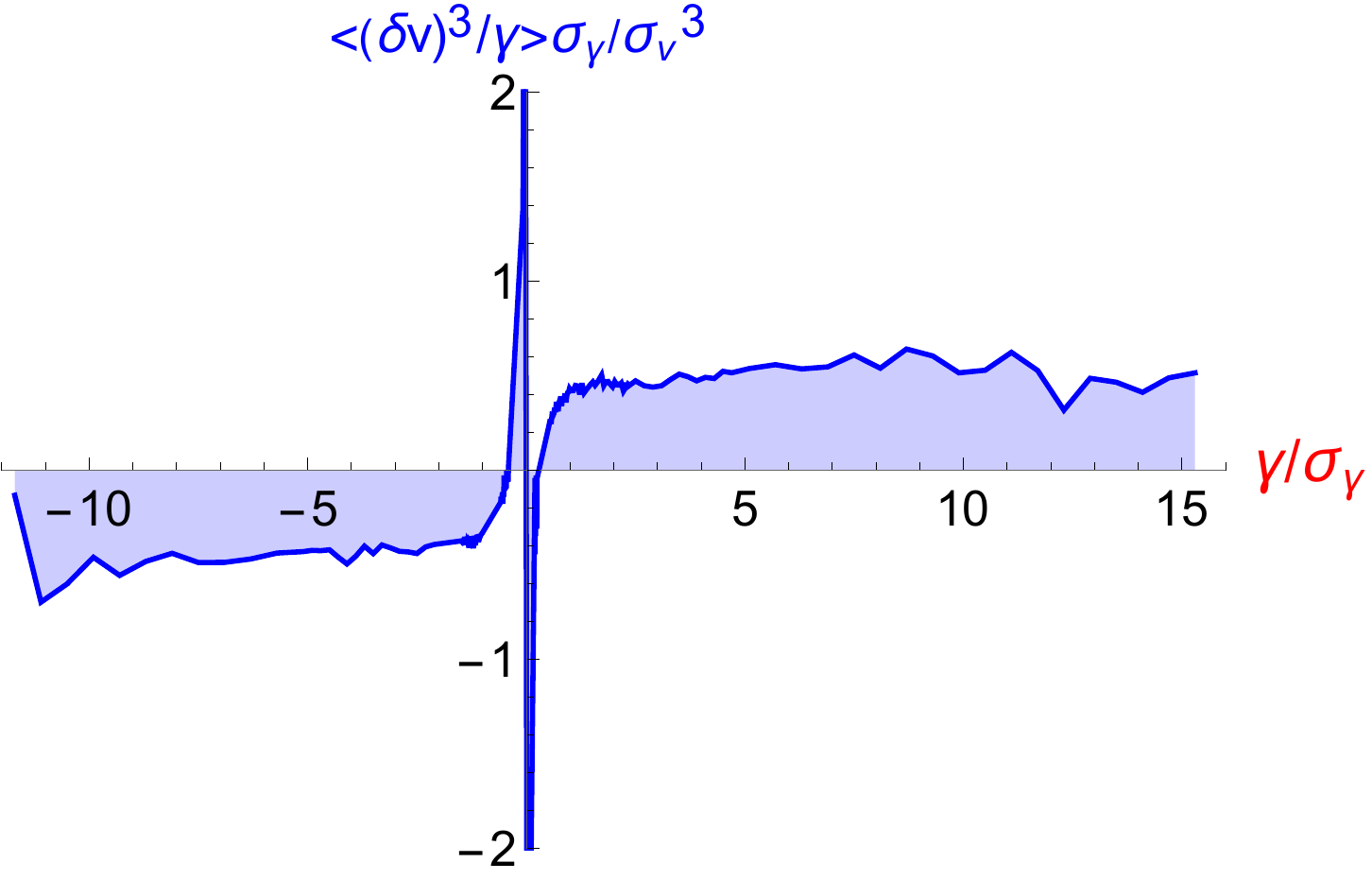}
(d)\includegraphics[height=2.0in]{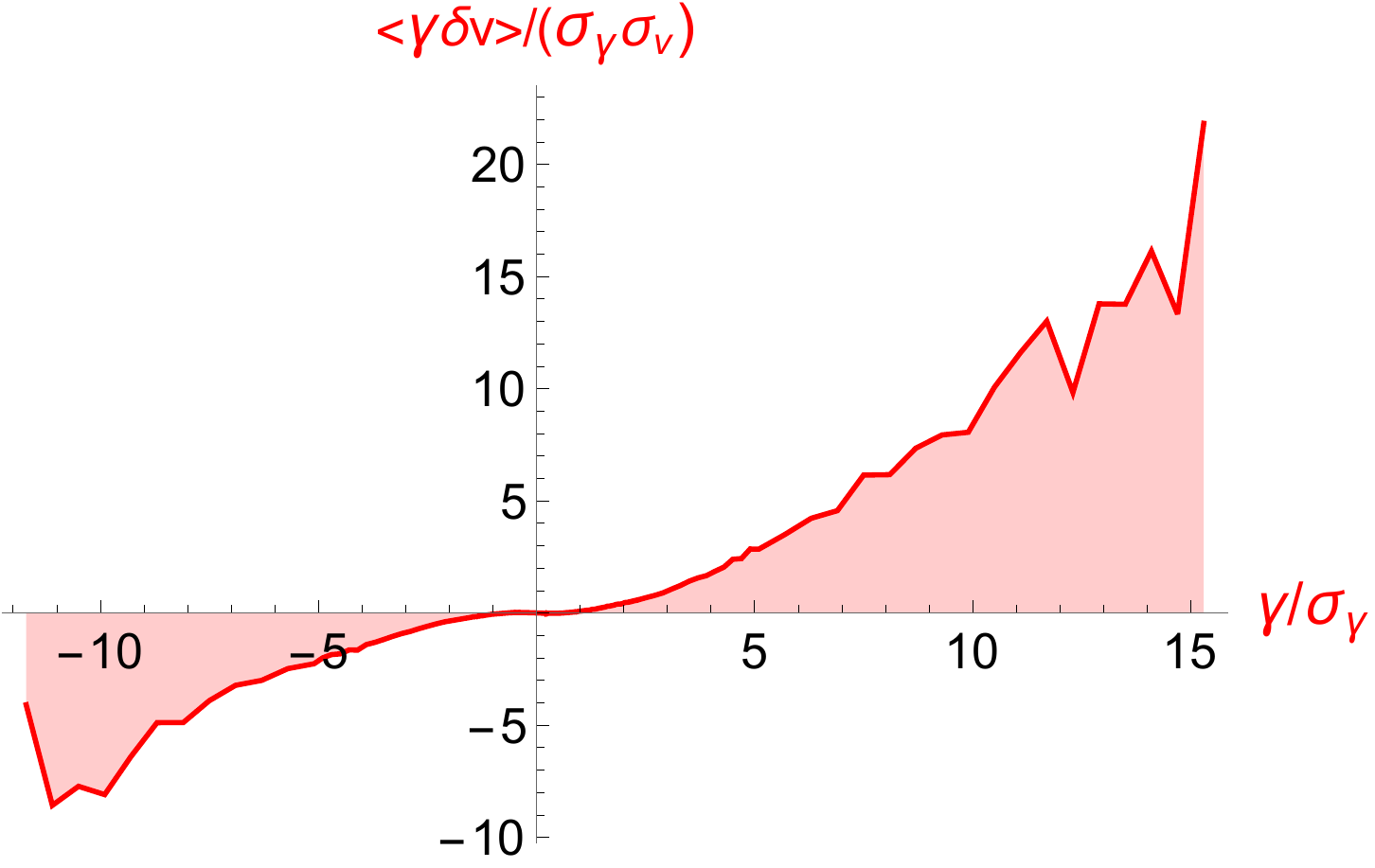}
  }
\caption{ Experimental  test to show the validity of relation (\ref{eq:cubv}), see text.The 
 conditional averages $ <(\delta v^{3}) >_{\gamma}$ , and $<\delta v>_{\gamma}$, plotted   in  (a) and (b) respectively, display a monotonic growth with respect to the acceleration $\gamma$, the linear behavior of curve (a) is emphasized by the  red dashed straight segments.  
The conditional ratio $<(\delta v)^{3} /\gamma>_{\gamma}$  in (c)  presents  a quasi-constant behavior for  $\gamma \gtrsim \sigma_{\gamma}$ in agreement with (\ref{eq:cubv}). In (d) the product  $<(\delta v)\gamma>_{\gamma}$  shows a clear non-constant behavior, but a strong growth with respect to $\gamma$, except in a small interval close to the origin where (\ref{eq:Kolm})  is approximately valid. All curves  are calculated via the  probability density of the set $[\delta v, \gamma] $   taken at the same time.  The velocity and acceleration are scaled to their respective standard deviation.}
\label{fig:test}
\end{figure}

To check which one of  relations (\ref{eq:cubv}) or (\ref{eq:Kolm}) is compatible with experimental data of Modane, we have plotted  $ <(\delta v)^{3} >/_{\gamma}$  and 
$<\delta v>_{\gamma}$ in Figs.\ref{fig:test}-(a)-(b). These  curves   assert that Kolmogorov scalings cannot fit  the experimental data in the domain of large and even moderate values of the acceleration, because the two conditional averages increases with $\gamma$ (therefore the product  $ < \gamma \delta v >$ cannot stay constant).  To evaluate more precisely the constant behavior of  the ratio $<u^{3}/\gamma>_{\gamma}$  (Leray scalings) and the non constant behavior of the product $ < u \gamma >_{\gamma} $ (Kolmogorov scalings) versus $\gamma$, we have plotted  these quantities in Figs. (c)-(d).
 It appears clearly that the Kolmogorov scaling leading to relation (\ref{eq:Kolm})  is incompatible with  the data of Modane, because the product $<\gamma u>_{\gamma} $  shown in Fig.(d) is definitely not constant in the  large interval of acceleration we have investigated. On the opposite  we find that the  ratio $<u^{3}/\gamma>_{\gamma}$ , shown in Fig.(c), displays a fairly constant behavior except in the domain of small acceleration (or  the order of  the standard deviation $\sigma_{\gamma}$). 

Now let us see if one may derive an order of magnitude of the local Reynolds number from our study.The ratio $<u^{3}>_{\gamma}/ \gamma$ is numerically equal to $\tilde{\Gamma}=1.33 10^{-3}$. From (\ref{eq:effcirc}), we have $ \tilde{\Gamma}/\nu = Re_{s} f(Re_{s})^{2}$, that gives  the following relation for  local Reynolds number, 
\begin{equation}
Re_{s} f(Re_{s})^{2} =85.6
\textrm{.} 
 \label{eq:Re2}
\end{equation}
  when taking the kinematic viscosity of air equal to $\nu \simeq 1.56 \;10^{-5}$, its value  at room temperature.  If the numerical function $f$ is of order unity, the  result in (\ref{eq:Re2}) gives a  local Reynolds number much smaller than the global Reynolds number  which was experimentally estimated as $Re=4.2\; 10^{5}$  (see above), but much larger than unity, the value it should take if the singularity were one of the ultimate outcomes of a Kolmogorov cascade stopped by viscosity effects. 
 
  In summary,  we have shown 
that  the linear relation (\ref{eq:cubv}) between  $u^3$  and the acceleration $\gamma$  is well verified experimentally for large acceleration values, see Fig.\ref{fig:test}-(a) and (c). This result is well explained within  the hypothesis of 
existence of Leray-type singularities in the flow. Differently  the relation (\ref{eq:Kolm})  is found to be invalid for acceleration values larger than its standard deviation, but can be approximately valid  in the domain $\gamma < \sigma_{\gamma}$,  namely close to small acceleration values, see Fig.\ref{fig:test}-(d). The non-validity of (\ref{eq:Kolm})  comes obviously from the fact that the original scaling by Kolmogorov,  $ u_r  \sim  (\epsilon r)^{1/3}$, even though it describes an average property of the velocity fluctuations cannot do it for large values of $u_r$: large values of $u_r$ would have to be linked to large distances, incompatible with large accelerations which concerns short distances. Kolmogorov scaling remains compatible however with the average properties of the fluctuations, but our linear relation between $u^3$ and $\gamma$ spans a rather wide range of values of $u$ and $\gamma$ and so could be hard to reconcile with a Kolmogorov scaling even on average. 

Note that  such a good fit  between (\ref{eq:cubv}) and experimental data is slightly unexpected because it implies that the pre-factor $\tilde{\Gamma}$  is  not changing much from one large fluctuation to the other. It could indicate that this Reynolds number dependent pre-factor is such that for some unknown reason it does not change appreciably in different realization of the singularity. 
Finally we may conclude from this study that besides the Kolmogorov cascade (already observed by using Modane's data \cite{modane}), it is quite probable that singularities exist  in the flow, and that they could  be of the Leray-form (\ref{eq:self}).

\section*{Acknowledgments}
 The authors are very grateful to B\'ereng\`ere Dubrulle,  Jean Ginibre, Christophe Josserand, Thierry Lehner and St\'ephane Popinet for very useful discussions.

{}

\end{document}